\documentclass[11pt,graphicx,twocolumn]{article}

\usepackage{amssymb,amsmath,amsfonts}
\usepackage[latin1]{inputenc}
\usepackage{graphicx}
\usepackage{graphics}
\usepackage{subfigure}
\usepackage{eepic,epsfig}
\begin{document}
\title{On the motion of a test particle around a global monopole in a modified gravity}
\author{\\ T. R. P. Caram\^es \thanks{E-mail: carames@if.uff.br}\\ Instituto
de F\'isica, Universidade Federal Fluminense, Niter\'oi-RJ, Brazil\\ E. R. 
Bezerra de Mello \thanks{E-mail: emello@fisica.ufpb.br}\\
Departamento de F\'{\i}sica-CCEN, Universidade Federal da Para\'{\i}ba\\
58.059-970, C. Postal 5.008, J. Pessoa, PB,  Brazil,\\ M. E. X. 
Guimar\~aes {\thanks{E-mail: emilia@if.uff.br}}\\ Instituto de F\'isica, 
Universidade Federal Fluminense, Niter\'oi-RJ, Brazil}
\maketitle
\begin{abstract}
In this paper we suggest an approach to analyse the motion of a test particle in the spacetime of a global monopole within a $f(R)$-like modified gravity. The field equations are written in a more simplified form in terms of $F(R)=\frac{df(R)}{dR}$. Since we are dealing with a spherically symmetric problem, $F(R)$ is expressed as a radial function ${\cal F}(r)\equiv{F(R(r))}$. So, the choice of a specific form for $f(R)$ will be equivalent to adopt an {\it Ansatz} for ${\cal F}(r)$. By choosing an explicit functional form for ${\cal F}(r)$ we obtain the weak field solutions for the metric tensor, compute the time-like geodesics and analyse the motion of a massive test particle. An interesting feature is an emerging attractive force exerted by the monopole on the particle. 
\end{abstract}
\maketitle
\section{Introduction}
\label{intro}

A global monopole is a kind of topological defect which arises in certain gauge theories due to the spontaneous symmetry breaking (SSB). Such SSB processes may have been undergone by our universe as a consequence of many phase transitions in the early times. Global monopoles arise in a theory exhibiting SSB of the global gauge group $O(3)$ to $U(1)$ \cite{kibble,vilenkin}.  The gravitational field of this object in the context of General Relativy (GR) was investigated by M. Barriola and A. Vilenkin \cite{barriola}. There, the authors have shown that the spacetime associated with this object is characterized by a non-trivial topology represented by a deficit solid angle and presents a non-vanishing scalar curvature. Later, Barros and Romero \cite{romero} analysed the gravitational field of a global monopole in the context of the Brans-Dicke theory of gravity considering the weak field approximation. There a comparison is made with the corresponding result obtained from GR.

In the late ten years, several types of modified theories of gravity have been suggested as possible alternatives to explain the late time cosmic speed-up experienced by our Universe, one of these theories are the so-called $f(R)$ theories of gravity \cite{Odintsov,Carrol,Fay}. Such theories avoid the Ostrogradski's instability which is commonly observed in general higher derivatives theories \cite{Ostr,Woo}. 

The gravitational field of a global monopole in such modified theory of gravity has been investigated in \cite{carames1}. There, we have found solutions for the metric tensor in the weak field approximation considering a specific {\it Ansatz} for the functional $f(R)$. The field equations are expressed in terms of the function $F(R)=\frac{df(R)}{dR}$, in a similar approach as developed in \cite{Mut} and \cite{carames}. Since the system under investigation has a spherical symmetry, it was possible to write $F(R)$ as a function of the radial coordinate, $r$, only. In the present paper we intend to follow the same procedure previously adopted by us, nevertheless going further and investigating a wider class of arbitrary $n$-degree polynomial functions ${\cal F}(r)$. We consider some approximations. For instance, the weak field limit and the modified gravity as a small correction on GR. 

In the scenario described above, we analyse the motion of a massive test particle in the spacetime of a global monopole from a classical point of view. The analysis of the classical motion of a massive test charged particle in the spacetime of a global monopole has been developed in \cite{Sobreira}, taking into accout the presence of the induced electrostatic self-interaction. A study of the motion of particles
in the field of a non-Abelian monopole in Einstein gravity was made in \cite{Jutta}.

This paper is organized as follows: In the section \ref{GM} we review briefly the model proposed by Barriola and Vilenkin for the global monopole in the context of GR. In the section \ref{Field} we introduce the modified theory of gravity and the field equations in the metric formalism, which is a crucial point to obtain in section \ref{FEs} the solution for the global monopole in $f(R)$ gravity. In the section \ref{Motion} we analyse the classical motion of a test particle in this spacetime.
Finally, we leave for section \ref{Conc} the most important concluding remarks about this paper.

\section{A Brief Review about Global Monopole}
\label{GM}

The model describing a global monopole is given by the following Lagrangian density \cite{barriola}:
\begin{equation}
\label{lagrangian}
{\cal L}=\frac{1}{2}\partial_{\mu}\phi^a \partial^{\mu}\phi^a-
\frac{1}{4} \lambda (\phi^{a}\phi^{a}-\eta^2)^2\ .
\end{equation}
Where $\phi^a$ is a self-coupling triplet of scalar fields and $a=1,2,3$. Plugging the energy-momentum tensor arising from (\ref{lagrangian}) into the Einstein equations, spherically symmetric solutions can be obtained, by adopting for the line element and the matter fields the {\it Ans\"atze} below:
\begin{eqnarray}
\label{sph}
ds^2=B(r)dt^2-A(r)dr^2-r^2d\Omega^2
\end{eqnarray}
and
\begin{equation}
\label{campo}
\phi^{a}=\eta h(r) \frac{x^{a}}{r} \ ,
\end{equation}
with $x^ax^a=r^2$ and $d\Omega^2=d\theta^2 + \sin^2\theta d\varphi^2$. Barriola and Vilenkin consider points far from the monopole's core where the energy-momentum tensor associated with the static matter field (\ref{campo}) can be approximated as
\begin{equation}
\label{emt}
T_\nu^\mu\approx{\rm diag}\left(\frac{\eta^2}{r^2},\frac{\eta^2}{r^2},0,0\right)\ .
\end{equation}
Taking these assumptions into account they have obtained the following solutions for the functions $B(r)$
and $A(r)$
\begin{eqnarray}
B = A^{-1}\approx 1 - 8\pi G\eta^2 - 2GM/r \ ,
\end{eqnarray}
being $M$ the mas of the monopole.

Furthermore, they have computed the light deflection performed by the gravitational field of the monopole and found
\begin{equation}
\delta \phi = 8 \pi G \eta^{2} l(d+l)^{-1}\ ,
\end{equation}
where $d$ and $l$ are the distances from the monopole to the observer and to the source, respectively. The light ray is constrained to lie at the equatorial plane $\theta= \frac{\pi}{2}$.

\section{The $f(R)$ Gravity in the Metric Formalism}
\label{Field}

In a $f(R)$ theory, the action associated with the gravitational field coupled to a matter field is given by:
\begin{equation}
\label{Act}
S=\frac1{2\kappa}\int d^{4}x\sqrt{-g}f(R)+{\cal S}_m \ ,
\end{equation}
where $f(R)$ is an analytical function of the Ricci scalar, $R$,
$\kappa=8\pi G$, being $G$ the Newton constant, and ${\cal S}_m$ corresponds to the action associated with the matter fields. By using the metric formalism, the field equations turn out to be:
\begin{eqnarray}
\label{FE}
F(R)R_{\mu\nu}&-&\frac{1}{2}f(R)g_{\mu\nu}-\nabla_{\mu}\nabla_{\nu}F(R)\nonumber\\&+&g_{\mu\nu}\Box F(R)=
\kappa T^{m}_{\mu\nu}\ ,
\end{eqnarray}
where $F(R)\equiv\frac{df(R)}{dR}$ and ${T}^{m}_{\mu\nu}$ is the standard minimally coupled energy-momentum tensor derived from the matter action, ${\cal S}_m $.

By taking the trace of equations (\ref{FE}) we get another one which shows us explicitly an emerging scalar degree of freedom in the modified gravity:
\begin{equation}
\label{trace}
F(R)R-2f(R)+3\Box F(R)=\kappa T^{m}\ .
\end{equation}
This expression allows us to write the function $f(R)$ in terms of its derivatives as follows
\begin{equation}
\label{fr}
f(R)=\frac{1}{2}\left(F(R)R+3\Box F(R)-\kappa T^{m}\right)\ .
\end{equation}
When (\ref{fr}) is plugged into (\ref{FE}) it is possible to express the field equations in terms of $F(R)$ as shown below:
\begin{align}
\label{feq}
&&F(R)R_{\mu\nu}-\nabla_{\mu}\nabla_{\nu}F(R)-\kappa \tilde{T}^{m}_{\mu\nu}=\nonumber\\&&\frac{g_{\mu\nu}}{4}\left[F(R)R-\Box F(R)-\kappa T^{m}\right] \ .
\end{align}
Of course, it is simpler to handle diferential equations involving $F(R)$ than to deal with them in terms of $f(R)$, since the former function is the first derivative of the latter one. So when this replacement is made the order of the field equations gets reduced by an unity. Other important reason for working out our approach looking always at $F(R)$ as a dynamical variable, instead of $f(R)$, is the new feature arising from equation (\ref{fr}). This equation says that a further scalar degree of freedom appears in in a $f(R)$ theory and, according to that equation, is exactly $F(R)$ which carries this information.  

From the expression (\ref{feq}) we can see that the combination below
\begin{align}
\label{A0}
	C_\mu=\frac{F(R)R_{\mu\mu}-\nabla_{\mu}\nabla_{\mu}F(R)-\kappa T^{m}_{\mu\mu}}{g_{\mu\mu}} \ ,
\end{align}
with fixed indices, is independent of the corresponding index. So, the following relation
\begin{equation}
\label{Ceq}
C_{\mu}-C_{\nu}=0\ ,
\end{equation}
holds for all $\mu$ and $\nu$.

\section{The Field Equations}
\label{FEs}

Substituting the energy-momentum tensor \eqref{emt} into (\ref{Ceq}), in \cite{carames1} 
we have derived the field equations associated with a global monopole 
system in modified theories of gravity. Considering for the metric tensor the {\it Ansatz}
given in (\ref{sph}) we have:
\begin{equation}
\label{Y1}
	2r{\cal F}^{''}-r{\cal F}'\left(\frac{B'}{B}+\frac{A'}{A}\right)-2{\cal F}\left(\frac{B'}{B}+\frac{A'}{A}\right)=0 \ ,
\end{equation}
and
\begin{eqnarray}
\label{Y2}
&-&4B+4AB-4rB\frac{{\cal F}'}{{\cal F}}+2r^{2}B'\frac{{\cal F}'}{{\cal F}} \nonumber\\&-&r^{2}B'\left(\frac{B'}{B} +\frac{A'}{A}\right)+2r^{2}B^{''}\nonumber\\&+&2Br\left(\frac{B'}{B}+\frac{A'}{A}\right)-\frac{4AB \kappa \eta^2}{\cal F}=0\ , \nonumber\\
\end{eqnarray}
where we have expressed $F(R)$ as $F(R(r))={\cal F}(r)$, since the problem we are analyzing has a radial symmetry, and primes mean derivative with respect to $r$.
Those equations may be simplified if we consider the following definition
\begin{equation}
\label{beta2}
\beta\equiv \frac{B'}{B}+\frac{A'}{A}\ ,
\end{equation}
which will allow us to write the field equations as follows
\begin{equation}
\label{beta}
\frac{\beta}{r}=\frac{{\cal F}''}{{\cal F}}-\frac{1}{2}\frac{{\cal F}'}{{\cal F}}\beta\ ,
\end{equation}
and
\begin{eqnarray}
\label{beta1}
&-&4B+4AB-4rB\frac{{\cal F}'}{{\cal F}}+2r^{2}B'\frac{{\cal F}'}{{\cal F}}+2r^{2}B^{''}\nonumber\\
&-&r^2B'\beta+2Br\beta-\frac{4AB\kappa \eta^2}{{\cal F}}=0\ .
\end{eqnarray}
As it has pointed out in \cite{carames1}, it is possible to verify that if one plugs $\omega=0$ (Brans-Dicke parameter) into the equation $(9)$ of reference \cite{romero} and replaces $\phi(r)$ by ${\cal F}(r)$, the equation (\ref{beta}) is recovered.
Which is a clear evidence of the equivalence of the $f(R)$ gravity in the metric formalism with a Brans-Dicke gravity possessing a parameter $\omega=0$, as previously discussed in \cite{faraoni}. Furthermore this fact offers us a deep ground to work with the field equations in terms of $F(R)$, since the results obtained by us can be compareable with those ones found via Brans-Dicke gravity. In this case, the role of the scalar degree of freedom in $f(R)$ gravity is played by the function $F(R)$.

In order to simplify our analysis possible, we shall assume some approximations. First of all, we choose to handle the field equations in the weak field regime, which means to consider that the components of the metric tensor are given by: $B(r)=1+b(r)$ and $A(r)=1+a(r)$ with $\left|b(r)\right|$ and $\left|a(r)\right|$ much smaller than unity. We also consider that the modified gravity is just slightly deviation from GR, i.e ${\cal F}(r)=1+\psi(r)$, with $|\psi(r)|<<1$. Furthermore, since $ G \eta^2\approx 10^{-5}$ in G.U.T., we may keep only terms linear in $G \eta^2$ and $\psi(r)$. By ruling out all the crossed terms of such small quantities and their derivatives, we get a linearized form for the original field equations   
\begin{equation}
\label{ap1}
\frac{\beta}{r}=\psi''
\end{equation}
and
\begin{eqnarray}
\label{ap2}
&&2a-2r\psi'+r\beta+r^2b''\nonumber\\&-&2\kappa \eta^2=0\ .
\end{eqnarray}

Closed results for equations (\ref{ap1}) and (\ref{ap2}) can be obtained by adopting for ${\cal F}(r)$ a specific {\it Ansatz} which means merely to solve them for a given $\psi(r)$ as it will be done in the next subsection. 

\subsection{The Solution for ${\cal F}(r)=1+\psi_0 r^n$}

Let us consider for the function $\psi(r)$ a power law-like {\it Ansatz}, namely $\psi(r)=\psi_0 r^n$, where $\psi_0$ is a constant parameter to be determined associated with the modification of the gravity. It is convenient to impose $\psi(r)$ as being regular at the origin which implies that $\psi(0)=0$. This assumption automatically rules out all the negative powers of $n$. Moreover, we assume that $\psi(r)$ is an analytical function of $r$, so that it admits a Taylor series expansion. The equations (\ref{ap1}) and (\ref{ap2}) for the aforementioned {\it Ansatz} read, respectively
\begin{equation}
\label{ap3}
a'+b'=n(n-1)\psi_{0}r^{n-1}
\end{equation}
and
\begin{eqnarray}
\label{ap4}
&&2a-2n\psi_{0}r^{n}+r(a'+b')+r^2b''\nonumber\\&&-2\kappa \eta^2=0\ .
\end{eqnarray}

In order to ensure that the metric is asymptotically flat as $\psi_0\rightarrow0$, the integration constant arising from (\ref{ap3}) is set to be zero, therefore a following relation between $a(r)$ and $b(r)$ is obtained:
\begin{equation}
\label{relab}
a+b=(n-1)\psi_0r^n\ .
\end{equation}
The relation above can be used in (\ref{ap4}) in order to express it in terms of $b(r)$
\begin{equation}
\label{eqb}
r^2b''-2b+(n+1)(n-2)\psi_0r^n-2\kappa \eta^2=0\ , 
\end{equation}
whose solution is
\begin{equation}
b(r)=\frac{c_1}{r}-c_2 r^2-\psi_0r^n-\kappa \eta^2\ ,
\end{equation}
where the integration constants $c_1$ and $c_2$ are defined, respectively, as $c_1=-2GM$ and $c_2=0$, to recover suitably the Newtonian potential present and due to the absence of a cosmological constant in the model we are analysing. Since $B(r)=1+b(r)$ we have:
\begin{equation}
\label{sol1}
B(r)=1-\frac{2GM}{r}-8\pi G\eta^2-\psi_0 r^n\ . 
\end{equation}
From equation (\ref{beta2}) we obtain
\begin{equation}
A(r)B(r)=a_0 e^{(n-1)\psi_0 r^n}\ ,
\end{equation}
where we set the integration constant $a_0$ to be unity in order to have a spacetime obeying the asymptotical flatness condition. Therefore,
\begin{equation}
\label{sol2}
A(r)=e^{(n-1)\psi_0 r^n}\left[1-\frac{2GM}{r}-8\pi G\eta^2-\psi_0 r^n\right]^{-1}. \
\end{equation}
Following the same reasoning of Barriola and Vilenkin, we drop out the mass term which is negligible at astrophysical scale, so we may write
\begin{equation}
\label{solb}
B(r)=1-8\pi G\eta^2-\psi_0 r^n\
\end{equation}
and
\begin{equation}
\label{sola}
A(r)=e^{(n-1)\psi_0 r^n}\left(1-8\pi G\eta^2-\psi_0 r^n\right)^{-1}\ .
\end{equation}
A binomial expansion could be applied in the solution (\ref{sola}) so that it would be expressed as
\begin{equation}
\label{solapp}
A(r)\approx 1+8\pi G\eta^2+n\psi_0 r^n\ .
\end{equation}

It is important to recall that our analysis is restricted to a specific range of the radial coordinate $r$, which is
\begin{equation}
\label{range}
\delta<r<\frac{1}{|\psi_0|^{\frac{1}{n}}}\ ,
\end{equation}
being $\delta\approx\left(\lambda \eta^{1/2} \right)^{-1}$ of the order of magnitude of the monopole's core. The reason for that, is because we have 
derived the field equations outside the monopole's core where $h\approx 1$; moreover, due to the weak field approximation we must have $|\psi_0\ r^n|<1$.

The solutions (\ref{solb}) and (\ref{solapp}) above allow us to compute the corresponding Ricci scalar
\begin{eqnarray}
\label{RicciM}
&&R=-\left[(n-1)(n-2)+2(n+2)\right]\psi_0 r^{n-2}\nonumber\\&-&\frac{16\pi G \eta^2}{r^2}\ ,
\end{eqnarray}
from which it is possible to determine the explicit form of $f(R)$. The procedure to be followed consists in inverting the equation above by expressing $r$ as a function of $R$, then plugging $r=r(R)$ into ${\cal F}(r)=1+\psi_0r^n$ and finally performing the integration of $F(R)$ which will give $f(R)$ plus an integration constant, which we may discard if no cosmological constant is taken into account in the theory. For any physically viable $f(R)$ theory under consideration, the following stability conditions must be fullfilled \cite{pogosian}-\cite{faraoni}:
\begin{itemize}
\item $\frac{d^2f(R)}{dR^2}>0$ (no tachyons);
\item $\frac{df(R)}{dR}>0$ (no ghosts);
\item $\lim_{R \rightarrow \infty} \frac{\Delta}{R}=0$ and $\lim_{R \rightarrow \infty}
\frac{d \Delta}{dR}=0$ (GR is recovered at early times),
\end{itemize}
where $\Delta=\Delta(R)$ is defined as $\Delta=f(R)-R$.

\subsubsection{The conformal relation with Barriola-Vilenkin metric}

It can be verified that the line element described by (\ref{solb}) and (\ref{solapp}) are conformally related to the global monopole solution in GR. Let us consider the transformation of coordinates below:
\begin{eqnarray}
B(r)=p(\bar{r})\left(1-8\pi G \eta^2\right)\ , \\
\label{transfA}
A(r)dr^2=p(\bar{r})\left(1+8 \pi G \eta^2\right)d\bar{r}^2\ , \\
r=p^{1/2}(\bar{r})\bar{r}\ ,
\label{trans}
\end{eqnarray}
where $p(\bar{r})$ is an arbitrary function of $\bar{r}$ to be determined and $p(\bar{r})=1+q(\bar{r})$ with $\left|q(\bar{r})\right|<1$.
Diferentiating equation (\ref{trans}) we have
\begin{equation}
\label{eqpq}
dr^2=\left(1+\bar{r}\frac{dq}{d\bar{r}}+q\right)d\bar{r}^{2}\ .
\end{equation}  

Substituting Eq. (\ref{eqpq}) into (\ref{transfA}) we obtain the following result for $q(\bar{r})$ (we keep only linear terms in $q(\bar{r})$, $\psi_0r$ and $G\eta^2$): 
\begin{equation}
q(\bar{r})=-\psi_0\bar{r}^{n}\ , 
\end{equation}
then
\begin{equation}
p(\bar{r})=1-\psi_0\bar{r}^{n}\ .
\end{equation}
Thus we can write the line element (\ref{sph}) in the coordinate $\bar{r}$ as follows
\begin{eqnarray}
\label{line1}
&&ds^2=\left(1-\psi_0\bar{r}^{n}\right)\left[\left(1-8\pi G \eta^2\right)dt^2\right. \nonumber\\
&&-\left.\left(1+8\pi G \eta^2\right)d\bar{r}^2\right. \nonumber\\
&&-\left.\bar{r}^2\left(d\theta^2+\sin^2\theta d\phi^2 \right)\right]\ .
\end{eqnarray}
If we rescale the time coordinate and redefine the radial coordinate as $r=\left(1+4 \pi G \eta^2 \right)\bar{r}$ we arrive at the line element below
\begin{eqnarray}
\label{line2}
ds^2&=&\left(1-\psi_0r^n\right)\left[dt^2-dr^2-\left(1-8 \pi G \eta^2 \right)\right.\nonumber\\&& \left.\times r^2\left(d\theta^2+\sin^2\theta d\phi^2 \right)\right]\ .
\end{eqnarray}
An important feature arises if we are interested in analyzing the deflection of light in this metric. As it is well known, the deflection angles are always preserved for two metrics related by a conformal transformation. Therefore, the deflection of a light ray by the monopole in the present modified gravity, considering weak field approximation, will be the same of that one previously obtained in Ref.\cite{barriola}.

\section{The Classical Motion of a Test Particle}\label{Motion}
In this section we analyse the classical motion of a test particle in the spacetime whose metric tensor has the form \eqref{sph} with components
given by (\ref{solb}) and \eqref{solapp}. From the corresponding line element we can define the Lagrangean of a test particle moving on this geometry as follows:
\begin{eqnarray}
\label{lagrangean}
\left(\frac{ds}{d\tau}\right)^{2}&=&2{\cal L}=B(r)\dot{t}^{2}-A(r)\dot{r}^{2}\nonumber\\&-&r^2\dot{\theta}^{2}-r^2\sin^2\theta\dot{\varphi}^{2}\ ,
\end{eqnarray}
where the dot means derivative with respect to $\tau$.

For orbits at equatorial plane, i.e., with $\theta=\frac{\pi}{2}$, the corresponding canonical momenta, $p_{\alpha}=\frac{\partial {\cal L}}{\partial \dot{x}^{\alpha}}$ are,
\begin{eqnarray}
\label{momenta}
p_{t}=E=B(r)\dot{t}\ ,\;\;\;\;\;\; p_{\theta}=-L_{\theta}=0\ ,\;\;\;\;\;\; \nonumber\\
p_{r}=-A(r)\dot{r}\;\;\;\;\;\;\ ,\;\;\;\;\;\;\ p_{\varphi}=-L_{\varphi}=-r^2\dot{\varphi}\ ,
\end{eqnarray}
where the constants of motion $E$ and $L_{\varphi}$ are interpreted, respectively, as the energy, and angular momentum, per unit mass, in $\varphi$-direction of the particle. 

For a massive particle the relation below holds
\begin{equation}
g_{\mu\nu}\frac{dx^{\mu}}{d\tau}\frac{dx^{\nu}}{d\tau}=1\ ,
\end{equation}
from which we obtain
\begin{equation}
\label{geodesic}
\left[1+(n-1)\psi_0r^{n}\right]\dot{r}^2+V_{\tiny eff}(r)=E^2\ .
\end{equation}
That shows a position-dependence of the particle's mass. Moreover, $V_{\tiny eff}(r)$ is the effective potential energy associated with the test particle and is defined as
\begin{equation}
\label{eff}
V_{\tiny eff}(r)=\left(1-8\pi G \eta^2-\psi_{0}r^n\right)\left(1+\frac{L^2}{r^2}\right)\ ,
\end{equation}
with $L\equiv L_{\varphi}$. 

In the next subsections, we shall analyze more detailedly the classical motion of the test particle.

\subsection{The tangential motion}
For this case it is necessary to require the stable circular motion conditions given by
\begin{itemize}
\item[(i)] $\dot{r}=0$\ ,

\item[(ii)] $\frac{\partial V_{\tiny eff}}{\partial r}=0$\ ,

\item[(iii)] $\frac{\partial^{2}V_{\tiny eff}}{\partial r}>0$\ .
\end{itemize}
The first and second conditions provide us the following system of equations:
\begin{equation}
\frac{E^{2}}{B}-\frac{L^{2}}{r^2}-1=0
\end{equation}
and
\begin{equation}
\frac{\partial}{\partial r}\left[A^{-1}\left(\frac{E^{2}}{B}-\frac{L^{2}}{r^2}-1\right)\right]=0\ .
\end{equation}
Whose solution for $E$ and $L$ are, respectively, given by:
\begin{equation}
\label{en}
E=B\sqrt{\frac{2}{2B-B'r}}
\end{equation}
and
\begin{equation}
\label{ang}
L=\sqrt{\frac{B'r^3}{2B-B'r}}\ .
\end{equation}

The angular velocity for any equatorial orbit is defined as
\begin{equation}
\Omega\equiv\frac{d\varphi}{dt}=\frac{d\varphi/d\tau}{dt/d\tau}\ .
\end{equation}
So, from the equations (\ref{en}) and (\ref{ang}) we obtain:
\begin{equation}
\Omega=\frac{B}{r^2}\frac{L}{E}=\sqrt{\frac{B'}{2r}}\ .
\end{equation}
It is well known that the velocity of the test particle along the $i$-direction is \cite{landau,frolov}:
\begin{equation}
v_{i}^{2}=-\frac{g_{ii}}{g_{00}}\frac{dx^i\ dx^i}{dt^2}\ .
\end{equation}
Thus, the tangential velocity $v_{\varphi}$ reads
\begin{equation}
v_{\varphi}^{2}=-\frac{g_{\varphi\varphi}}{g_{00}}\left(\frac{d\varphi}{dt}\right)^2=-\frac{g_{\varphi\varphi}}{g_{00}}\Omega^2\ .
\end{equation}
By working out the expression above and keeping only linear terms in $G\eta^2$ and $\psi_0$ we obtain
\begin{equation}
\label{orbit}
v_{\varphi}=\sqrt{-\frac{1}{2}n\psi_0r^n}\ .
\end{equation}
The expression above says that circular orbits will be physically allowed only if $\psi_0<0$. Hence, such equation can be rewritten as
\begin{equation}
\label{orbit1}
v_{\varphi}=\sqrt{\frac{1}{2}n\left|\psi_0\right|r^n}\ ,
\end{equation}
which shows us that in the case under consideration, with the mass of the monopole's core considered as negligible, the circular motion of the test particle is consequence of the modification parameter of the gravity, $\psi_0$. 

\subsection{The emerging extra force}
One important physical property observed when the global monopole is analyzed within a $f(R)$ gravity is the gravitational force exerted by the monopole on a massive particle moving nearby. It is also important to emphasize that this feature is absent in GR, since the Barriola-Vilenkin 
monopole has a metric with $g_{00}$=const. which means that no gravitational force exist. On the other hand, based on our previous result, we have found 
\begin{align}
g_{00}=1-8\pi G \eta^2-\psi_{0}r^n \ , 	
\end{align}
which gives rise to a radial force acting on the particle for $\psi_{0}\neq 0$. 

As it is well known, the motion of the test particle in a weak gravitational field is described by the equation \cite{landau}
\begin{equation}
\label{force0}
\ddot{x}^{i}=-\frac{1}{2}\frac{\partial h_{00}}{\partial x^{i}}\ ,
\end{equation}
being $h_{00}$ the deviation form the unity in $g_{00}$. In order to obtain the force given by (\ref{force0}), let us express the metric (\ref{sph}) in galilean coordinates, $g_{\mu\nu}=\eta_{\mu\nu}+h_{\mu\nu}$, by performing the transformation below 
\begin{equation}
\label{transf1}
t=\left(1-4\pi G\eta^2\right){\cal T}\ ,
\end{equation}
and

\begin{equation}
\label{transf2}
r=\left(1-4\pi G\eta^2-\frac{4\pi G\eta^2\ln \left(\left|\psi_0\right|{\cal R}^{n}\right)}{n}\right){\cal R}\ ,
\end{equation}
which will imply

\begin{eqnarray}
\label{elemento1}
ds^2&=&\left(1-8\pi G \eta^2-\psi_{0}{\cal R}^n\right)d{\cal T}^2\nonumber\\
&-&\left[1-16\pi G \eta^2-\psi_0{\cal R}^{n}\right.\nonumber\\&-&\left.\frac{8\pi G \eta^2 \ln \left(\left|\psi_0\right|{\cal R}^{n}\right)}{n}\right]\times \nonumber\\
&\times&\left(d{\cal R}^2+{\cal R}^2d\Omega^2\right)\ ,\nonumber\\
\end{eqnarray}
where ${\cal R}=\sqrt{x^{2}+y^{2}+z^{2}}$\ . Hence, by working out (\ref{force0}) for (\ref{elemento1}) we obtain
\begin{equation}
\ddot{x}^{i}=\frac{n\psi_{0}{\cal R}^{n-1}}{2}\frac{x^{i}}{{\cal R}}\ ,
\end{equation}
whose corresponding force experienced by an unit mass particle is
\begin{equation}
\label{force}
\stackrel{\rightarrow}{F}=\frac{n\psi_{0}{\cal R}^{n-1}}{2}\hat{{\cal R}}\ .
\end{equation}
Because of the negativeness required for $\psi_0$ such force will have an attractive nature which may be evidenced by expressing (\ref{force}) as
\begin{equation}
\label{force1}
\stackrel{\rightarrow}{F}=-\frac{n\left|\psi_{0}\right|{\cal R}^{n-1}}{2}\hat{{\cal R}}\ .
\end{equation}
Therefore, as a consequence of modifying GR we have obtained an extra and attractive force exerted by the global monopole on a massive test particle. 

It is useful to investigate what are the possible motions that the particle can perform. In order to do that, let us express the effective potential in terms of a set of dimensionless variables defined as follows
\begin{eqnarray}
&&\tilde{r}\equiv(\eta \lambda^{1/2})r\ ,\;\;|\tilde{\psi_{0}}|\equiv\left|\psi_0\right| 
\left(\eta \lambda^{1/2}\right)^{-n}\ ,\nonumber\\ &&\tilde{L^{2}}\equiv\ L^2\left(\eta^{2}\lambda\right)\ ,\;\;\alpha^2\equiv1-8\pi G \eta^2\ .
\end{eqnarray}
With these changes the range of validity (\ref{range}) is redefined as
\begin{equation}
\label{range1}
1<\tilde{r}<\frac{1}{{|\tilde{\psi_{0}}|}^{\frac{1}{n}}}\ .
\end{equation}
A remarkable feature which is important to be emphasized is that the range of validity presented above is shorter the higher is the value of $n$. We mean, the size of the region in space in which some important change will be observable due to the modification of gravity tends to zero as higher powers of $\psi(r)$ are considered.
 
In terms of those new variables the effective potential (\ref{eff}) may be written as
\begin{equation}
\label{eff1}
V_{\tiny eff}(\tilde{r})=\alpha^2+|\tilde{\psi_{0}}|\tilde{r}^{n}+\alpha^2\frac{\tilde{L^{2}}}{\tilde{r^2}}-\tilde{\psi_0}\tilde{L^{2}}\tilde{r}^{n-2}\ .
\end{equation}

By assigning numerical values for the parameters $\alpha^2$, $\tilde{\psi_{0}}$, $\tilde{L}$ and $n$, present in the equation above, we can analyse the behavior of the possible profiles of the effective potential. Let us make plots of $V_{\tiny eff}(\tilde{r})$ against $\tilde{r}$ for different values of $\tilde{\psi_{0}}$ and then observe how the effective potential behaves for especific values of $n$, namely $n=1,2$ and $3$, respectively.

\subsubsection{Plot for $n=1$}
A plot of $V_{\tiny eff}(\tilde{r})$ against $\tilde{r}$ is exhibited in the Fig $1$. We have sketched the graph by assigning specific values for $\alpha^2$ and $\tilde{\psi_{0}}$. In this plot two distinct configurations of effective potentials are presented: In the one represented by solid line, compatible with the context of modified gravity, we can see that the attractive force traps a test particle moving with energy greater than the minimum potential. In the second one, represented by dashed line, the parameter responsible for modifying the gravity $\tilde{\psi_0}$ is set equal to zero and the potential has no extremal. So, in the latter, the particle will not be trapped by the monopole, as predicted by Barriola-Vilenkin since this monopole does not exert any gravitational force on the massive particles moving nearby.

\begin{figure}[!htb]
\begin{center}
\includegraphics[width=7.5cm,height=7.0cm]{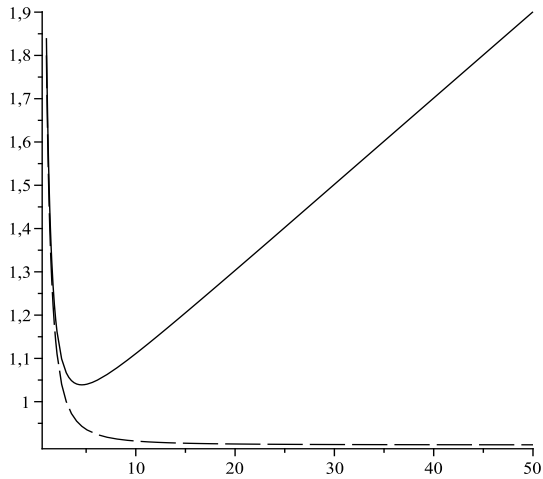}
\end{center}
{\caption{\footnotesize Plot of the effective potential $V_{\tiny eff}$ vs $\tilde{r}$ for two different ,set of values of $\alpha$, $\tilde{\psi_0}$ and $\tilde{L}$. The solid line shows a minimum of the effective potential for $\alpha^2=0.9$, $\tilde{\psi_{0}}=-0.02$ and $\tilde{L}=1$, which means that a particle moving with energy $E^{2}>V_{\tiny eff}(r_{\tiny min})$ will be trapped by the monopole. On the other hand, the dashed line expresses the corresponding effective potential within GR, with $\alpha^2=0.9$, $\tilde{\psi_{0}}=0$ and $\tilde{L}=1$. In the latter case, the potential has no minimum and the particle cannot be trapped by the monopole as expected from the result of Barriola-Vilenkin.}}
\label{f5}
\end{figure} 

\subsubsection{Further plots for $n=2$ and $3$}
The next examples we intend to analyse are that ones concerning to $n=2$ and $3$. The plots $V_{\tiny eff}$ vs $\tilde{r}$ shown below represent the profiles of the effective potential (\ref{eff1}) for each one of those values of $n$. The numerical values of $\tilde{\psi_{0}}$ adopted in each case were $\tilde{\psi_{0}}=-0.02$. The different scales of the horizontal axis in each plot remarks the dependence of the range of validity (\ref{range1}) upon the value adopted for $n$. 

\begin{figure}[h]
\begin{minipage}[]{0.60\linewidth}
\includegraphics[width=\linewidth]{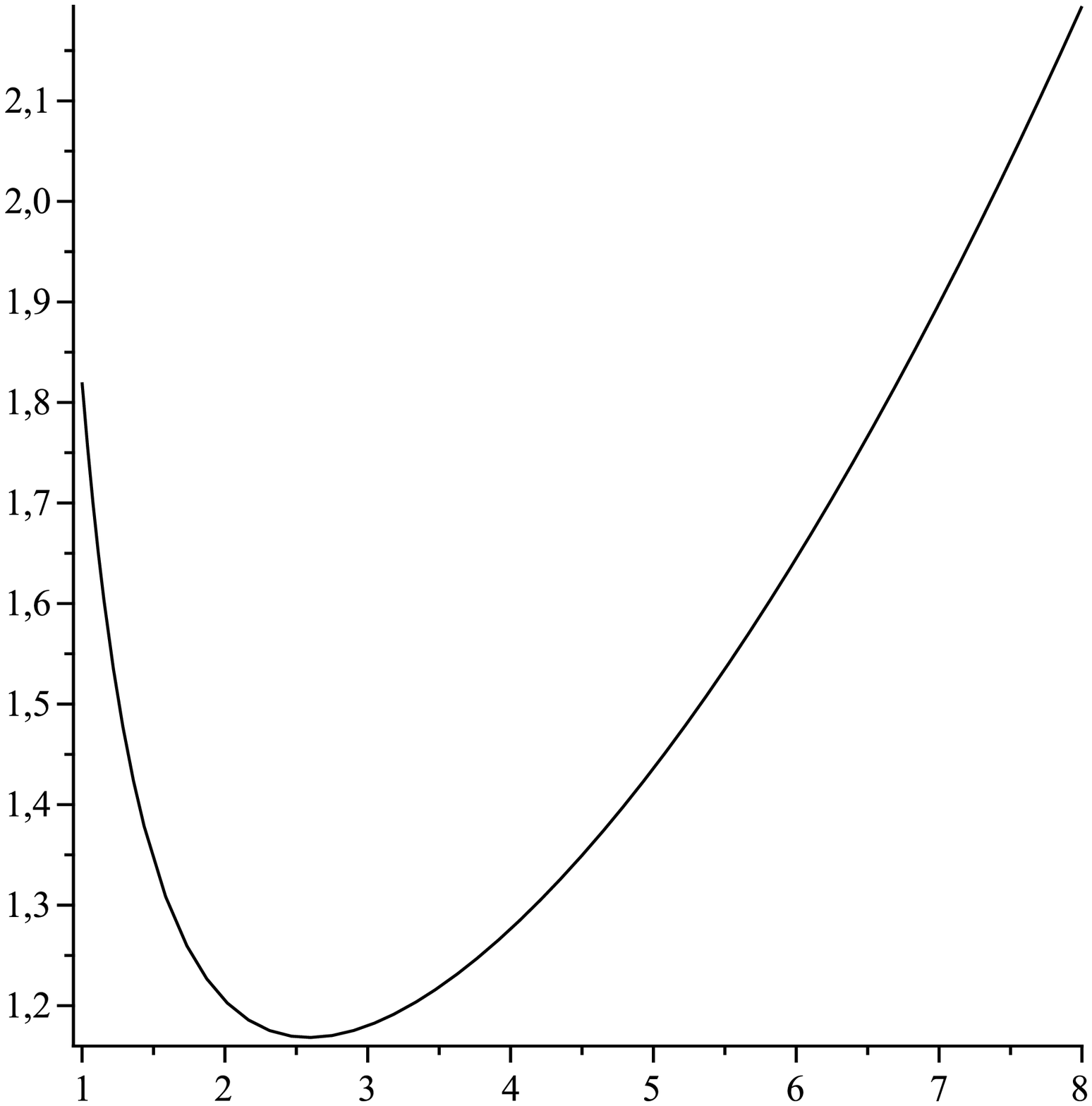}
\caption{\footnotesize Case $n=2$. For such case the range of validity is $1<\tilde{r}<7.07$, so it is reasonable to leave the radial coordinate to vary between $1$ and $8$ as it is well expressed in the plot.}
\label{fig1}
\end{minipage} \hfill
\begin{minipage}[h]{0.60\linewidth}
\includegraphics[width=\linewidth]{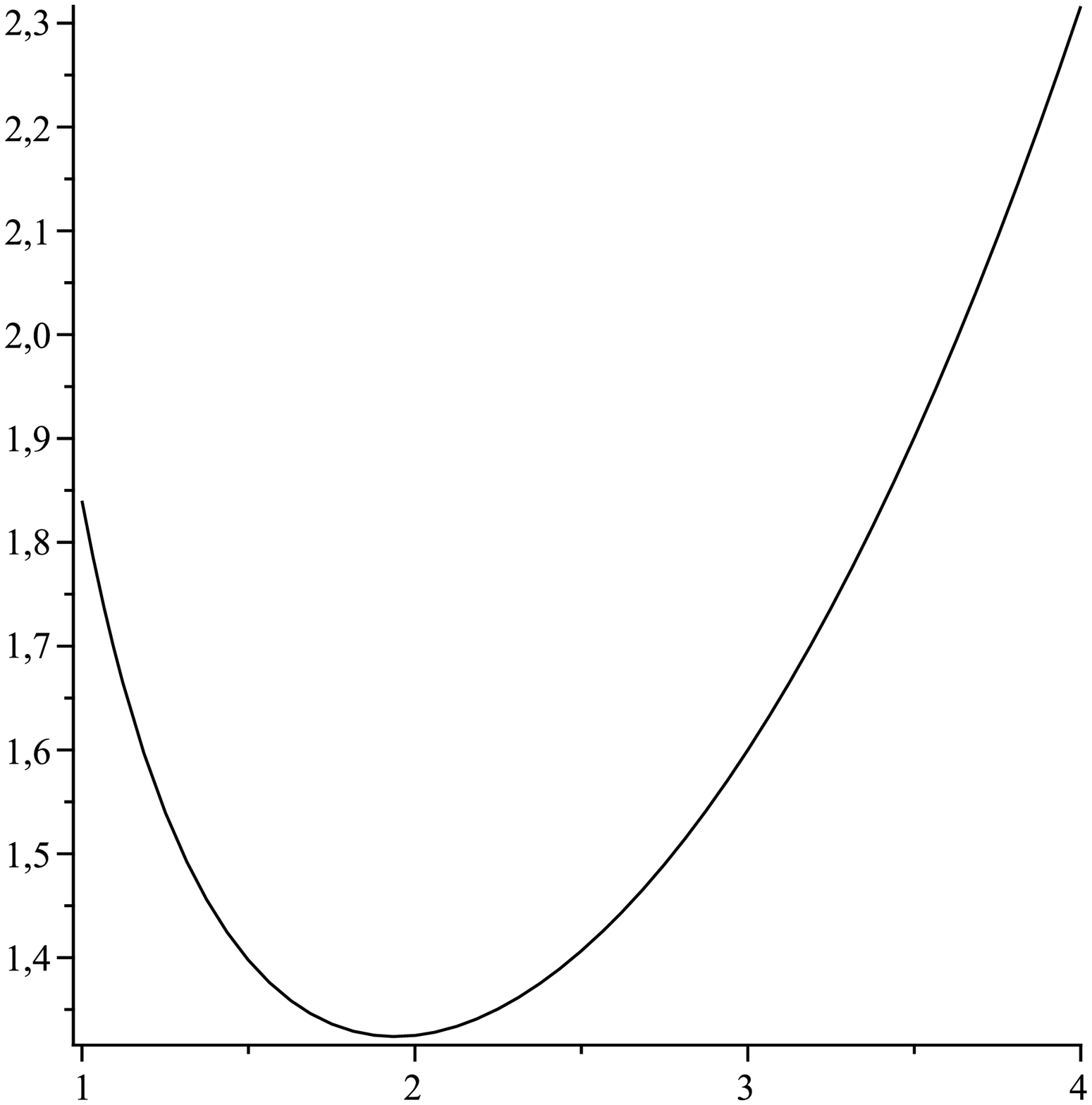}
\caption{\footnotesize Case $n=3$. Now the range of validity is $1<\tilde{r}<3.68$, which suggests the variation shown in the horizontal axis, namely between $1$ and $4$.}
\end{minipage}
\end{figure}

As we can see, for increasing values of $n$, the graphs trend to keep similar profiles, always having a well-defined minimum which is necessary for the arising of the extra force. Since the force (\ref{force}) is attractive, it will be possible to observe a trapping of the test particle by the Global Monopole if the particle moves with energy greater than the minimum potential. In each case sketched above there will be two turning points for the motion of the particle. This trapping is a feature absent within GR, being an exclusive consequence of this modification in gravity we did. 

\section{Conclusion}\label{Conc}

In this paper we have analyzed the classical motion of a massive test particle in the gravitational field of a global monopole in the $f(R)$ gravity scenario in the metric formalism. By this formalism, this modified gravity contains a massive scalar degree of freedom in addition to the familiar massless graviton and it turns out to be equivalent to a Brans-Dicke theory. In order to simplify our analysis we have considered solutions in the weak field approximation, which implied that we are considering a theory that is as a small correction on GR. The latter condition was explicitly considered by assuming ${\cal{F}}(r)=1+\psi_0r^n$, for $|\psi_0r^n|<1$, being $n$ an positive integer number.

Following the above mentioned approach, the solutions found by us correspond to small corrections on $g_{00}$ and $g_{11}$ components of the metric tensor, only. Being these new components given in (\ref{sol1}) and (\ref{sol2}), respectively. From the results obtained, we can observe that, a small correction on the Ricci scalar, given now by (\ref{RicciM}), takes place. Moreover, we have also verified that these solutions are conformally related to that one previously obtained by M. Barriola and A. Vilenkin as shown in (\ref{line2}), what ensures us that the deflection
of light in these two spacetimes will be the same.

As to the classical motion, we have derived the differential equation, given by Eq. \eqref{geodesic}, which allows us to analyse separately the tangential and radial motions. As our main conclusion, we have observed that the presence of the parameter $\psi_0$ is essential to provide stable circular orbits for the particle. This is a new feature, because in the context of general relativity and discarding the mass term in the metric tensor, there is no Newtoninan gravitational potential produced by a global monopole, consequently, no stable orbits for the particle is  possible.

An important result we have found was an emerging force arising within the modified gravity, which depends on the parameter $\psi_0$ and also on the values attributed to $n$. Such force, as verified by us, has an attractive nature and becomes stronger for larger values of $n$. This is a new effect arising in the modified theory, being absent within Barriola-Vilenkin model. In the last section we have sketched graphs for the effective potential for specific values of $n$, namely $n=1,\;2$ and $3$ and observed that their behavior are similar, in the sense that they always present attractive emerging forces. 

\section{Acknowledgments}

TRPC thanks CAPES for financial support and S. Jor\'as for fruitful discussions and Orahcio Sousa for the computational aid. ERBM and MEXG thank Conselho Nacional de Desenvolvimento Cient\'\i fico e Tecnol\'ogico (CNPq) for partial financial support.


\begin{thebibliography}{100}    

\bibitem{kibble} T. W. Kibble, {\it J. Phys. A: Math. Gen.} {\bf 9}, 1378 (1976).

\bibitem{vilenkin} A. Vilenkin and E. P. Shellard, {\it Cosmic String and Other
Topological Defects} (Cambridge University Press, Cambridge, 1994).

\bibitem{barriola} M. Barriola and A. Vilenkin, {\it Phys. Rev. Lett.} {\bf 63}, 341 (1989).

\bibitem{romero} A. Barros and C. Romero, {\it Phys. Rev. D} {\bf 56}, 6688 (1997).

\bibitem{Odintsov} S. Nojiri and S. D. Odintsov, Phys. Rev. D {\bf 68}, 125312 (2203); S. Nojiri and S. D. Odintsov, Phys. Lett. B {\bf 657}, 238 (2007).

\bibitem{Carrol} S. M. Carrol, V. Duvvuri, M. Trodden and M. S. Turner,
{\it Phys. Rev. D} {\bf 70}, 043528 (2004).

\bibitem{Fay} S. Fay, R. Tavakol and S. Tsujikawa, {\it Phys. Rev. D} {\bf 74}, 063509 (2007).

\bibitem{Ostr} M. Ostrogradski, {\it Mem. Ac. St. Petersbourg VI} {\bf 4}, 385 (1850).

\bibitem{Woo} R. P. Woodard, {\it Lect. Notes. Phys.} {\bf 720}, 403 (2007).

\bibitem{carames1} T. R. P. Caram\^es, E. R. B. de Mello and M. E. X. Guimar\~aes,  Int. J. of Modern Physics: Conference Series, v. 03, p. 446-454, (2011) .

\bibitem{Mut} T. Multamaki and I. Vilja, {\it Phys. Rev. D} {\bf 74}, 064022 (2006).

\bibitem{carames} T. R. P. Caram\^es and E. R. B. de Mello,
{\it Eur. Phys. J. C} {\bf 64}, 113  (2009).

\bibitem{Sobreira} A. A. Sobreira and E. R. Bezerra de Mello, {\it Grav.
Cosm.} {\bf 5}, 177 (1999).

\bibitem{Jutta} V. Kagramanova, J. Kunz and C. Laemmerzahl  {\it Gen. Rel. Grav.} {\bf 40}: 1249-1278, (2008).

\bibitem{pogosian} L. Pogosian and A. Silvestri, {\it Phys. Rev. D} {\bf 77}, 023503 (2008).

\bibitem{joras} V. Miranda, S. E. Jor\'as, I. Waga and M. Quartin,
{\it Phys. Rev. Lett.} {\bf 102}, 221101 (2009); S. E. Jor\'as, {\it Int. J. Modern Physics A} {\bf 26}, 3730, (2011). 

\bibitem{dolgov} A. D. Dolgov and M. Kawasaki, {\it Phys. Lett. B} {\bf 573}, 1 (2003).

\bibitem{faraoni} V. Faraoni, {\it Phys. Rev. D} {\bf 74}, 104017 (2006).

\bibitem{landau} L. D. Landau and E. M. Lifshitz, {\it The Classical Theory of Fields}, Vol. 2, 4th edition, (Butterworth-Heinemann, 1975).

\bibitem{frolov} V. P. Frolov and I. D. Novikov, {\it Black Hole Physics: Basic Concepts and New Developments}, (Kluwer Academic Publishers, 1989).

\end{thebibliography}
\end{document}